\documentclass[manuscript]{aastex}

\shorttitle{Statistics of Polarization Measurements}
\shortauthors{Strohmayer et al.}

\begin{document}


\title{On the Statistical Analysis of X-ray Polarization Measurements}


\author{T. E. Strohmayer and T. R. Kallman}
\affil{X-ray Astrophysics Lab, Astrophysics Science Division, NASA's Goddard 
Space Flight Center, Greenbelt, MD 20771}

%





\begin{abstract}
In many polarimetry applications, including observations in the X-ray
band, the measurement of a polarization signal can be reduced to the
detection and quantification of a deviation from uniformity of a
distribution of measured angles of the form $A + B \cos^2 ( \phi -
\phi_0 ) \; (0 < \phi < \pi)$.  We explore the statistics of such
polarization measurements using Monte Carlo simulations and $\chi^2$
fitting methods.  We compare our results to those derived using the
traditional probability density used to characterize polarization
measurements and quantify how they deviate as the intrinsic modulation
amplitude grows.  We derive relations for the number of counts
required to reach a given detection level (parameterized by $\beta$
the ``number of $\sigma's$'' of the measurement) appropriate for
measuring the modulation amplitude $a$ by itself (single interesting
parameter case) or jointly with the position angle $\phi$ (two
interesting parameters case).  We show that for the former case when
the intrinsic amplitude is equal to the well known minimum detectable
polarization (MDP) it is, on average, detected at the $3\sigma$
level. For the latter case, when one requires a joint measurement at
the same confidence level, then more counts are needed than that
required to achieve the MDP level. This additional factor is
amplitude-dependent, but is $\approx 2.2$ for intrinsic amplitudes
less than about $20\%$. It decreases slowly with amplitude and is
$\approx 1.8$ when the amplitude is $50\%$. We find that the position
angle uncertainty at $1\sigma$ confidence is well described by the
relation $\sigma_{\phi} = 28.5 ({\rm degrees}) / \beta$.

\end{abstract}





\section{Introduction}

Emission and scattering processes thought to be important in many
astrophysical X-ray sources are likely to impart specific polarization
signatures, but to date there have only been a few positive detections
of polarization from cosmic X-ray sources, largely due to sensitivity
limitations.  Some of the earliest and highest precision measurements
were made with the OSO-8 Bragg reflection polarimeter (Kestenbaum et
al. 1976; Weisskopf et al. 1976), and include high significance
measurements of the linear polarization properties of the Crab nebula
in several energy bands (Weisskopf et al. 1978).

More recently, observations with the {\it INTEGRAL} spectrometer SPI
and imager IBIS have exploited the polarization dependence of Compton
scattering to infer the linear polarization properties of the Crab at
$\gamma$-ray energies (Dean et al. 2008; Forot et al. 2009).  These
results indicate that the $> 200$ keV flux from the Crab nebula is
highly polarized ($\approx 50 \%$), with a position angle consistent
with the pulsar rotation axis.

In the last few years the development of micropattern gas detectors
has enabled the capability to directly image the charge tracks
produced by photoelectrons, thus enabling use of photoelectric
absorption in a detection gas as a direct probe of X-ray polarization
(Costa et al. 2001; Black et al. 2004; 2010).  It is likely that such
technology will be employed in the not-too-distant future to
sensitively explore the polarization properties of many classes of
astrophysical X-ray sources for the first time.


In this paper we explore the question of how one detects, measures,
and characterizes a polarization signal with a photoelectric
polarimeter.  The remainder of this paper is organized as follows; in
\S 2 we outline the basic problem of detecting a modulation in a
distribution of angles, and we describe the angle distributions used
throughout the paper.  In \S 3 we briefly outline the probability
distribution relevant to such polarization measurements. In \S 4 we
describe our Monte Carlo simulations and present our results. We
conclude with a short summary in \S 5.

\section{Statement of the Problem}

The angular distribution of photoelectrons ejected by a linearly
polarized beam of photons (X-rays) is proportional to $\sin^2 (\theta
) \cos^2 (\phi ) / (1-\beta\cos(\theta ) )^4$ (see, for example, Costa
et al. 2001), where $\theta$ is the emission angle measured from the
direction of the incident photon ($0 < \theta < \pi$), and $\phi$ is
the azimuthal angle measured relative to the polarization vector of
the incident photon (see Figure 1 for the basic geometry applicable to
a photoelectric polarimeter).  In most practical situations the angle
$\theta$ is not inferred directly, but the electron charge track
projected into the plane orthogonal to the direction of the photon
(the plane defined by $\theta = 90$ deg) is imaged and thus $\phi$ can
be estimated for each detected photon. The angle $\phi$ is measured
around the line of sight to the target of interest and can be
referenced to, for example, local North on the sky.  In principle
$\phi$ can be measured in the range from $0 - 2\pi$ (0 - 360 degrees),
however, due to the two-fold symmetry of the $\cos^2 ( \phi )$
dependence of the angular distribution above, it is sufficient to
consider distributions over the range of angles from 0 to $\pi$ (0 -
180 degrees).

The presence of a significant linear polarization component is then
evident in the distribution of azimuthal angles. For example, an
unpolarized photon flux will produce a uniform distribution in the
angle $\phi$, whereas a linear polarization component produces a
distribution peaked at a particular azimuthal angle, $\phi_0$. We
emphasize that such a polarimeter does not require rotation of the
detector to achieve its sensitivity. Rather, because a photoelectron
is ejected preferentially--though with a quantum mechanical
probability distribution--along the polarization vector of the photon,
measurement of the distribution of the azimuthal angles of
photoelectrons provides a {\it direct} probe of the linear
polarization properties of the source.

Due to the $\cos^2 (\phi)$ dependence of the angular distribution of
photoelectrons, the observed number density of photon events with
measured angle between $\phi$ and $\phi+d\phi$, which we denote as
$S(\phi) d\phi$, can be expressed in the form  (see, for example,
Pacciani et al. 2003; Costa et al. 2001),
\begin{equation}
S (\phi ) = A + B \cos^2 \left ( \phi - \phi_0 \right ) .
\end{equation}

The total intensity of photons is then just the integral,
$\int_{0}^{\pi} S(\phi) d\phi$, which in this case can be readily
shown to be $(A + B/2)\pi$.  From equation 1 the unpolarized
(unmodulated) component of the intensity is evidently $A\pi$, and
since the total intensity can be expressed as the sum of the polarized
and unpolarized intensities, the polarized intensity is then simply
$B\pi/2$.  This angular distribution has an amplitude of modulation
defined as, $a \equiv (S_{max} - S_{min}) / (S_{max} + S_{min}) = B /
(2A + B)$, and a position angle (the angle at which the distribution
has a maximum) given by $\phi_0$, thus, the detection of polarization
can be reduced to a statistical detection of a modulation in the
distribution of angles, $\phi$.  Such a distribution is often referred
to as a modulation curve.

Now, the amplitude of modulation $a$ is not in general equal to the
source polarization amplitude, $a_p$, because a detector is not a
perfect analyzer and will not provide an exact measurement of the true
photoelectron angle $\phi$. That is, individual angle estimates will
have some uncertainty associated with them and these will produce a
uniform (unmodulated) component to the measured distribution even in
the case of a 100\% polarized beam. This ``lossiness'' of the angle
estimates is quantified in terms of the so-called detector modulation,
$\mu$, which is the modulation amplitude produced in the detector by a
100\% polarized X-ray beam. 

In general, $\mu$ can depend on a number of factors, including the
energy of the incident photons, and the composition and pressure of
the absorbing gas, among others (see Paciani et al. 2003, Bellazini et
al. 2003). In the absence of background, the amplitude of polarization
is just $a_p = a / \mu$. In general, $a_p$ is larger than the measured
amplitude of modulation, $a$, because as noted above, detectors are
not 100\% efficient, and some of the intrinsic polarization amplitude
is smeared out.

An X-ray source's linear polarization characteristics can thus be
described by the modulation amplitude, $a$, and position angle, $\phi$
(in the range from 0 - $\pi$), that would be produced in a particular
detector system.  In making an observation a total of $N$ photons are
observed for each of which an angle $\phi_i$ is estimated for the
ejected photoelectron in the range from $0 - \pi$ (0 - 180 deg).  One
can then create a histogram, or modulation curve, of the number of
events (photons), $n_j$, in each of $M$ position angle bins, where the
bin size (in degrees) is $\Delta\phi = 180/M$, and then perform
least-squares ($\chi^2$) fitting to estimate both $a$, $\phi$, and
their $1\sigma$ uncertainties. We call this a measurement of
polarization. The counts in any position angle bin, $n_j$, are
independent, Poisson-distributed random variables whose expectation
value is determined by $S(\phi)$.  It is the statistics of such
measurements that we explore in the remainder of this paper. In
effect, what is measured is the modulation amplitude, $a$, and it is
the knowledge of the detector system, expressed in terms of $\mu$,
that enables this to be converted to a source polarization amplitude
via the expression $a_p = a / \mu$.


With the help of some trigonometric identities equation 1 can be
re-written as,
\begin{equation}
S(\phi) = A + B\left [ \frac{1}{2} \left ( 1 +  \cos (2\phi) \cos (2\phi_0) +
\sin (2\phi) \sin (2\phi_0) \right ) \right ] \; ,
\end{equation}
which is equivalent to,
\begin{equation}
S(\phi) = \left (A + \frac{B}{2} \right ) + \left ( \frac{B}{2}\sin
(2\phi_0) \right ) \sin (2\phi) + \left ( \frac{B}{2}\cos (2\phi_0) \right ) 
\cos (2\phi) \;.
\end{equation}
If we define $I = (A + B/2)$, $Q = (B/2)\sin (2\phi_0)$, and $U =
(B/2)\cos (2\phi_0)$, this gives the so-called Stokes decomposition
(in this case specific to linear polarization),

\begin{equation}
S(\phi ) = I + Q \sin \left ( 2 \phi \right ) + U \cos \left ( 2 \phi 
\right ) ,
\end{equation}
where $I$, $Q$ and $U$ are the well-known Stokes parameters.  The
modulation amplitude, $a = B/(2A + B)$, can now be written as $a =
(Q^2 + U^2 )^{1/2} / I$, and by dividing $Q$ by $U$ we can express the
position angle as $\phi_0 = 1/2 \tan^{-1} ( Q / U )$.  It is also
straightforward to show that in this form the unpolarized intensity is
given by $A\pi = ( I - (Q^2 + U^2)^{1/2} )\pi$, and the polarized
intensity $(B/2)\pi = (Q^2 + U^2)^{1/2} \pi$.  The sum of these gives
the total intensity, $I\pi = (A + B/2)\pi$.

We note that the form of equation 4 is familiar as a partial Fourier
series, with the Stokes parameters as the Fourier coefficients. 
We can argue that they are Gaussian distributed random variables to good
approximation as long as a reasonably large number of position angle
bins are used to construct the modulation curves.  To see this we
sketch an example of how one of the Stokes parameters can be expressed
as a Fourier coefficient.  If we multiply both sides of equation (4)
by $\cos (2\phi)$ and integrate we obtain,
\begin{equation}
\int_0^{\pi}S(\phi)\cos (2\phi) d\phi = I \int_0^{\pi} \cos (2\phi) d\phi
+ U\int_0^{\pi} \cos^2 (2\phi) d\phi + Q\int_0^{\pi} \cos (2\phi) \sin
(2\phi) d\phi \;.
\end{equation}
Because of the angular integrals only the 2nd term on the right-hand
side of equation (5) is non-zero, leading to the following expression
for $U$,
\begin{equation}
U = \frac{2}{\pi} \int_0^{\pi} S(\phi) \cos (2\phi) d\phi \; .
\end{equation}
Now, this integral can be approximated as a sum over all bins in the
measured (or simulated) modulation curve with $S(\phi_j) = n_j$, the
number of photon events in each bin.  Since the $n_j$ are random,
Poisson distributed values their sum will approach a normal
distribution as M (the number of bins in the modulation curve) becomes
large enough via the central limit theorem.  While this ensures that
the distributions will tend toward the normal distribution, it does
not necessarily enforce statistical independence of the parameters,
but this can be tested carefully with Monte Carlo simulations. Below
we show that under such circumstances the Stokes parameters are not
completely independent, but that in the limit of small amplitudes the
approximation of independence is a very good one. Indeed, we show that
while $Q$ and $U$ are statistically independent, both $Q$ and $U$ are
correlated with $I$ in an amplitude-dependent manner. Moreover, we
show that the distributions of $Q$ and $U$ are characterized by a
variance that decreases with increasing amplitude.

\section{Probability Distribution} 

Assuming the independent and normally distributed properties of the
Stokes parameters are satisfied, previous studies have shown that the
joint probability distribution for a measurement of linear
polarization characterized by amplitude, $a$, and position angle,
$\phi$ is given by
\begin{equation}
P (N, a, a_0, \phi, \phi_0) = \frac{Na}{4\pi} \exp \left ( -\frac{N}{4} 
(a^2 + a_0^2 - 2a a_0 \cos 2((\phi - \phi_0)) ) \right ) ,
\end{equation}
where $a$, $a_0$, $\phi$, $\phi_0$, and $N$ are the measured
amplitude, the true amplitude, the measured position angle, the true
position angle, and the number of detected photons, respectively (see,
for example, Simmons \& Stewart 1985; Vaillancourt 2006; Weisskopf,
Elsner \& O'Dell 2010).  In the case of no intrinsic polarization,
$a_0 = 0$, the distribution simplifies substantially to
\begin{equation}
P (N, a) =  \frac{Na}{4\pi}  \exp \left ( -\frac{N}{4} a^2 \right ) ,
\end{equation}
and this expression can be readily integrated to find the probability
of measuring an amplitude, $a$, if there is no intrinsic
polarization. The amplitude that has a 1\% chance of being measured is
referred to as the {\it minimum detectable amplitude} (MDA), and it is
relatively straightforward to show that $MDA = 4.29 / \sqrt N$.  The
polarization amplitude that would produce this modulation amplitude in
a particular detector system is called the {\it minimum detectable
polarization} (MDP), and based on the discussion above is just $MDP =
MDA / \mu = 4.29 / (\mu\sqrt N)$.

Furthermore, if we are not concerned with the position angle, $\phi$,
then we can integrate equation (3) over angles and obtain the
distribution,
\begin{equation}
P(N, a, a_0)=\int_{-\pi/2}^{\pi/2}P(a,\phi)d\phi
=\frac{a}{\sigma^2}exp(-\frac{(a^2+a_0^2)}{2\sigma^2})
I_0(\frac{aa_0}{\sigma^2})
\label{eqn1}
\end{equation}
\noindent where $I_0$ is the modified Bessel function of order zero,
and $\sigma^2 = 2/N$.  This distribution is known as the Rice
distribution (Rice 1945) and it reflects the fact that the amplitude
is always a positive quantity, and so the distribution must go to zero
for $a=0$.  An important property of this distribution concerns the
second moment, which is related to the width, and is given by:
$<a^2>=a_0^2+4/N$ which shows that the distribution width increases
with $a_0$.  

The Rice distribution has relevance to a number of other research
fields, including various signal processing applications (Abdi et
al. 2001), magnetic resonance imaging (Sijbers et al. 1998) and radar
signal analysis (Nilsson \& Glisson 1980; Marzetta 1995).  It is a
common goal in many of these applications to estimate the amplitude
parameter, $a$, of the Rice distribution from observed data.  We note
that the X-ray polarization case discussed here has similarities to
these other applications but is more general in the sense that one is
often interested in estimating both the amplitude, $a$, as well as the
position angle, $\phi$. It is beyond the scope of this paper to
explore and compare a wide range of different parameter estimation
methods, rather, here we focus on the case of least squares fitting
using $\chi^2$ methods. 


\section{Monte Carlo Simulations}

We emphasize that the probability distribution given in the previous
section is only rigorously correct under the assumption that the
Stokes parameters are normally distributed and independent. Here we
begin by exploring the accuracy of that assumption.  The procedure of
generating modulation curves based on a particular choice of $S(\phi)$
is quite amenable to simulation with Monte Carlo techniques, and here
we present results of such simulations, both in the case with $a_0 =
0$ and $a_0 > 0$, and essentially map-out the exact probability
density numerically.  For now we ignore background considerations and
work only in terms of modulation amplitudes, that is, the following
results can be considered applicable to any detector system,
regardless of the $\mu$ value that characterizes its polarization
sensitivity (see Elsner, O'Dell \& Weisskopf 2012 for further
discussion on the effect of backgrounds).

The simulations proceed as follows, for a given set of true parameter
values, $a_0$ and $\phi_0$, we compute a number, $M_{sim}$, of
simulated data sets. The total number of events in a particular
realization is Poisson distributed with a mean of $N$ photons. For the
case of $a_0 = 0$ the distribution of angles, $\phi$, is uniform,
which can be readily simulated with a random number generator that
produces uniform deviates.  When $a_0 > 0$ we sample random angles
from the true distributions by the so-called transformation method. We
first compute the cumulative distribution of $S (\phi)$, then draw
uniform deviates, $x$, and find the corresponding value of $\phi(x)$
in the cumulative distribution. We use the root finder {\it ZBRENT}
implemented in {\it IDL} to solve for $\phi$ given $x$. We can thus
draw a specific number of random events, $N$, from any true
distribution.

For each simulated data set we bin the resulting angles to form a
modulation curve and for each the Stokes form of the distribution
(equation 4) is fitted to determine best-fit values for $Q$, $U$, and
$I$, and their $1\sigma$ uncertainties, $\sigma_Q$, $\sigma_U$, and
$\sigma_I$, respectively.  We can then use the fitted Stokes
parameters to express the results in terms of $a$ and $\phi$ using the
expressions defined above.  The $1\sigma$ uncertainties, $\sigma_a$
and $\sigma_{\phi}$, can also be estimated by standard error
propagation methods, which yields,
\begin{equation}
\sigma_a^2 = a^2 \left ( \frac{Q^2 \sigma_Q^2}{(Q^2 + U^2)^2} + 
\frac{U^2 \sigma_U^2}{(Q^2 + U^2)^2} + \frac{\sigma_I^2}{I^2} 
\right )  ,
\end{equation}
and 
\begin{equation}
\sigma_{\phi} \; ({\rm deg})= (180/\pi) \left ( \sqrt(x^2) / (
2(1+x^2) ) \right ) \left ( (\sigma_Q / Q)^2 + (\sigma_U / U)^2 \right
)^{1/2} ,
\end{equation}
where, $x = Q / U$ , and $\sigma_I$, $\sigma_Q$, and $\sigma_U$ are
the standard $1\sigma$ uncertainties on the fitted quantities, $I$,
$Q$ and $U$. The procedure can then be repeated with different values
of $N$. All the Monte Carlo simulations described here were done using
{\it IDL}, and uniform deviates were obtained with {\it IDL's} random
number generator {\it randomu}. We use a least-squares fitting routine
developed within {\it IDL} that is based on MINPACK-1 (Markwardt
2009).  In all of the least squares fits all parameters are allowed to
vary.

The basic procedure is illustrated in Figure 2, which shows a
simulated modulation curve (solid histogram with error bars) and best
fitting model (smooth black curve). The fitted Stokes parameter
components are also indicated and labelled. This example is a single
statistical realization of 15,000 events in $M=16$ angle bins from the
distribution $S(\phi) = 10 + 5 cos^2 (\phi - 30 {\rm(deg)})$. The
best-fit values and $1\sigma$ standard errors from the fit for $I$,
$Q$, and $U$ are $936.94 \pm 7.65$, $148.29 \pm 10.76$, and $96.84 \pm
10.76$, respectively.  The inferred amplitude and position angle are
$a = (Q^2 + U^2)^{1/2} / I = 0.189 \pm 0.0115$ and $\phi_0 = (1/2)
tan^{-1} (Q/U) = 28.43 \pm 1.74$ deg, and are entirely consistent with
the true values $a = 5/(20 + 5) = 0.2$, and $\phi_0 = 30$ deg.

From the results of many such simulations one can compute the
distributions of the fitted parameters, $I$, $Q$ and $U$. In doing
this we confirm that these distributions are Gaussian to very good
accuracy. Examples are given in Figure 3, which shows the resulting
distributions of $I$, $Q$ and $U$ (with their means subtracted) from
10,000 realizations using the same angular distribution as used to
produce Figure 2, but with $M = 40$. Since the mean-subtracted
distributions of $Q$ and $U$ were consistent with each other, we fit a
Gaussian model to the summed distribution. The best-fitting Gaussian
models are also plotted in Figure 3.

We next explored simulations using a wide range of intrinsic
amplitudes, $a_0$, and searched for correlations amongst the resulting
mean-subtracted distributions. From this analysis it is evident that
the parameters $Q$ and $U$ are statistically independent for any
intrinsic amplitude.  That is, a scatter plot of $Q - <Q>$ versus $U -
<U>$ is circularly symmetric about the origin. However, we find that
both $Q$ and $U$ are not independent of $I$, indeed they are
correlated in a manner which is proportional to the intrinsic
amplitude.  Figure 4 shows a pair of scatter plots computed from a
simulation using the angular distribution $S(\phi) = 7 + 16 cos^2
(\phi - 55 {\rm (deg)})$, which has a large modulation amplitude of
$16/30 = 0.533$. Plotted are the mean-subtracted best-fit values of
$I$ versus $Q$ (black symbols), and $Q$ versus $U$ (red symbols) for
100,000 realizations.  These simulations were computed with a mean
number of counts $N = 30,000$.  A positive correlation is evident in
the $I$ versus $Q$ points (black symbols), but the $Q$ versus $U$
distribution appears circularly symmetric, that is, uncorrelated.  To
further quantify this we also computed the linear correlation
coefficients as a function of amplitude for a set of simulations. The
results are shown in Figure 5 where the coefficients for $Q$ versus
$U$, $Q$ versus $I$, and $U$ versus $I$ are denoted by the ``error
bar,'' ``square','' and ``$\times$'' symbols, respectively.  This
clearly demonstrates that $Q$ and $U$ are uncorrelated, but that $Q$
and $U$ are both correlated with $I$, with the magnitude of the
correlation coefficient increasing with amplitude.

In addition, we find that the variance of the distributions of $Q$ and
$U$ depends on the intrinsic amplitude.  To quantify this effect we
computed the fractional change in the standard deviation, $(\sigma -
\sigma_0)/\sigma_0$, for each fitted parameter ($I$, $Q$ and $U$) from
a large suite of simulations.  Because the variance of the
distributions in $Q$ and $U$ are equal, as a practical matter we
compute the average of the two. Here, $\sigma$ is the standard
deviation of each fitted parameter computed from simulations, and
$\sigma_0^2 = N/M^2$ and $2N/M^2$ for the distributions in $I$, and
$Q$ and $U$, respectively. These are the expected variances in the
limit that the intrinsic amplitude is small and equation (7) is exact.
The results are shown in Figure 6, where the values for $I$ and the
average of the $Q$ and $U$ distributions are denoted by the
``diamond'' and ``square'' symbols, respectively.  Each pair of points
here was computed from 400,000 realizations. From Figure 6 it is
evident that the variance of $I$ is independent of the amplitude and
is consistent with $\sigma_0^2 = N/M^2$, but that the variance in both
$Q$ and $U$ is systematically smaller than $\sigma_0$ in an
amplitude-dependent manner.  The inset panel of Figure 6 shows a
zoom-in of the low amplitude portion of the plot, and demonstrates
that the change is quite small for low amplitudes. Even at an
amplitude of $30\%$ the percentage decrease in the standard deviation
is only about $1\%$.  However, above $50\%$ the deviation increases
sharply, and approaches $20\%$ as the amplitude reaches unity.  A
polynomial in powers of the amplitude (up to $a^4$) provides a good
fit to the values for the $Q$ and $U$ distributions from these
simulations, and this functional form is the solid curve in Figure 6.
It has the form $(\sigma - \sigma_0)/\sigma_0 = P_0 - P_1 a_0 - P_2
a_0^2 - P_3 a_0^3 - P_4 a_0^4$. The best fitting coefficients for
$P_0$ through $P_4$ are, $-1.79 \times 10^{-3}$, $-0.0451$, $0.40$,
$-0.594$, and $0.458$, respectively.

The above results demonstrate that for the measurement of modulation
curves appropriate to X-ray polarimetric data the Stokes parameters
are independent, gaussian random variables to good approximation as
long as the intrinsic amplitude is not too large. Our results have
quantified how accurate the approximation is as a function of the
modulation amplitude.  For example, even at an intrinsic amplitude of
$30\%$ the distributions of $Q$ and $U$ are only about $1\%$ narrower
than the same distributions at zero amplitude.  Thus, for many
applications use of equation (7) is warranted, however, for cases
where large modulation amplitudes are expected, or very precise
results required, then some caution should be exercised in its use. 

\subsection{Results with $a_0 = 0$: Minimum Detectable Amplitude (MDA)}

For the case of $a_0 = 0$ we can compute the distribution of best-fit
amplitudes, $a$, for a large number of simulations. From this
distribution we can then estimate the value $a_{1\%}$ that has a 1\%
chance probability of being measured. This is just the familiar MDA
described above.  Figure 7 shows a comparison of the results from such
simulations (blue square symbols) versus the analytic expression given
above, $4.29 / \sqrt N$ (solid line). We computed $M_{sim} = 10,000$
simulations for these results, and we used $M = 16$ position angle
bins for the modulation curves. Figure 7 shows that the $a_0 = 0$
simulations are in good agreement with the analytic result, as we
would expect in this case since as we showed above equation (7) is
exact in the limit as $a_0 = 0$. This gives us further confidence that
our simulation procedures are correct.

\subsection{Results with $a_0 > 0$: Detection of Polarization}

We can now explore a number of issues with regard to detection of
polarization.  For example, how many counts are needed to measure a
modulation amplitude to a particular precision, and for a given
precision in the amplitude measurement, how accurately can the
position angle be measured?

To address these questions we perform additional simulations using
true distributions with specified amplitudes and position angles. For
the illustrative examples below we used two different amplitudes, both
with $\phi_0 = 0$, however, we have explored many different cases and
all the results summarized here are independent of the particular
values of $a_0$ or $\phi_0$ used. The example distributions described
below are; $S (\phi) = 10 + 1 \cos^2 (\phi - 0 )$ and $S (\phi) = 10 +
2 \cos^2 (\phi - 0)$. These examples correspond to intrinsic
modulation amplitudes, $a_0$, of $1/21$ and $2/22$, respectively. For
a specific detector these would correspond to polarization amplitudes,
$a_p = 1/(21 \mu )$, and $2/(22 \mu )$. We emphasize that for these
modest amplitudes use of equation (7) to describe the probability
density should be a very good approximation.

To determine the number of counts, $N$, needed to reach a polarization
sensitivity given by the MDP value we require that MDP $= 4.29 / \mu
\sqrt N = a_p = a/ \mu$, and thus $a = 4.29 / \sqrt N$. Note that here
$\mu$ cancels out, and to achieve MDP values at these amplitudes
requires $(21^2)*4.29^2 = 8116.21$, and $(22/2)^2*(4.29)^2 = 2226.90$
counts, respectively. Note that this may seem a trivial point, but
it's important to keep the terminology as precise as possible. Below
we will compare the counts needed to reach a certain MDP, $N_{mdp}$,
and the counts needed to measure the same amplitude of polarization to
a certain precision, $N$. While $N$ and $N_{mdp}$ will individually
depend on $\mu$, the ratio, $N / N_{mdp}$, cannot depend on $\mu$,
that is, it is independent of the detector system employed.

With $a_0 > 0$ we compute for different values of $N$ a large number,
$M_{sim}$, of realizations, to each of which is fitted the Stokes
distribution (equation 4) to determine the best fit values of $I$,
$Q$, and $U$ (and thus $a$ and $\phi$), and their $1\sigma$
uncertainties (as illustrated in Figure 2). We thus simulate the
distributions of $a$ and $\phi$ for different $N$.

\subsection{Single Parameter Confidence Regions}

Here we present the results of our simulations in several ways. First,
we used the results from $M_{sim} = 1,000$ realizations for different
values of $N$ to find the mean values of $a$, and $\phi$ and their
68\% confidence ranges. These are derived for each parameter
independently of the other, that is, they are 1-dimensional (single
parameter) ranges. For example, if one were interested in asking the
question, is a source (or population of sources) polarized, without
regard to the particular position angle of the electric vector, then
the 1-d distribution would be appropriate. We explore the joint,
2-dimensional distributions below in \S 4.4.

Figure 8 shows an example simulated distribution of measured
amplitudes, $a$, computed with $N = 20,000$ events and $a_0 = 2/22$
(with $\phi_0 = 0$). The left panel shows the differential (binned)
distribution, and the right panel shows the same results expressed as
a cumulative distribution.  We also computed similar distributions for
the measured position angle, $\phi$.

Procedurally we use the estimated cumulative distributions to identify
the mean of each distribution (ie., for both $a$ and $\phi$), and the
two values $a_{max}$ and $a_{min}$ that enclose 68\% of the
distribution.  This then gives the $1\sigma$ uncertainty,
$\sigma_{a,1d} = (a_{max} - a_{min})/2$ (we also compute the
corresponding values for the $\phi$ distribution).  We can then
compute the quantity $\beta_{1d} = a_{mean} / \sigma_{a,1d}$, which
can be thought of as the ``number of sigmas" of the measurement. We
now explore the behavior of several quantities as a function of
$\beta_{1d}$.

The first is the ratio $N / N_{mdp}$ (see Figure 9), where $N$ is
simply the number of events (photons) simulated (ie. the number of
observed counts in the modulation curve), and $N_{mdp} = 4.29^2 /
a_0^2$ is just the number of counts that would be required to reach an
MDA equal to the true amplitude, $a_0$.  This ratio can be thought of
as the additional observing time required to measure the true
amplitude to a given significance compared to the time needed to reach
an MDA equivalent to the true amplitude. How might we expect this
ratio to depend on $\beta$?  From equation (10) and the fact that
$\sigma_Q = \sigma_U$ it is straightforward to show that $\sigma^2_a /
a^2 \equiv \beta^{-2} \approx 2/(N a^2)$ (there is strict equality in
the limit of small $a$). We can then substitute $a^2 = a_{MDP}^2 =
(4.29)^2 / N_{MDP}$, and a little arithmetic then demonstrates that
$N/N_{MDP} = \beta^2 (2 / 4.29^2)$.  The black diamond symbols in
Figure 9 show the results of simulations using the example
distributions described above with $M_{sim} = 1,000$ for different
values of $N$, and that the expected $\beta^2$ dependence is a good
match to the simulations.

One can also present the results in a slightly different but
complementary way.  In the above simulations we have essentially
carried out many simulated observations and computed the distribution
of ``observed" values of the amplitude and position angle, but in
reality an observer may not have the luxury of making such a large
number of independent observations of a particular source. All any
observer can do is to observe some number, $N$, of photons from the
source, construct a modulation curve and fit it as we have done in the
simulations described above. From this procedure we obtain four
quantities, $a$, $\sigma_a$, $\phi$, and $\sigma_{\phi}$. This
constitutes a measurement of the polarization parameters, or more
simply, a measurement of polarization.  Here, the uncertainties,
$\sigma_a$, and $\sigma_{\phi}$, are obtained for a specified
confidence level and number of degrees of freedom. For example, for
the $1\sigma$ (68.3\%), single parameter confidence ranges we would
find the change in each parameter that produces a $\Delta\chi^2 = 1$
(while allowing the other parameter to vary in the fitting
procedure). For a 2-dimensional, joint confidence region at the same
level of confidence (68.3\%) we would find the $\Delta\chi^2 = 2.3$
contour in the $a$ - $\phi$ plane (see, for example, Lampton et
al. 1976).

For the diamond symbols in Figure 9 above we used the mean values and
$1\sigma$ uncertainties derived from the distributions computed from
many simulated observations to obtain $\beta_{1d}$, however, one can
also use the ``measured" quantities from each simulated observation to
compute $\beta_{obs,1d} = a / \sigma_{a,1d}$. One can then plot the
observed quantities for each simulated observation, where now $N_{mdp}
(a)$ is computed using the best-fit value for $a$ and the formula $a =
4.29 / N_{mdp}^{1/2}$.  We have done this and show the results in
Figure 9 with the colored symbols. That is, the colored points are
these ``measured" values from individual simulations. The red symbols
were computed with $N=10,000$, $a_0 = 2/22$, $\phi_0=0$ (deg), the
blue used $N=20,000$, $a_0 = 2/22$, $\phi_0=0$ (deg), and the green
with $N=24,000$, $a_0 = 3/25$, and $\phi_0=22.5$ (deg). We see that
the distribution of ``measured" points falls along the same relations
as that deduced from the simulated distributions of $a$, and $\phi$,
as indeed they should since they are sampling the same distributions.
This way of presenting the results of the simulations makes a more
direct connection with actual polarimetric observations, as we only
plot quantities that one would obtain directly from a single
observation. To the extent that actual observations are dominated only
by Poisson counting noise and for which the background is small, then
they must fall along the relations followed by the simulated
observations shown in Figure 6.  Indeed, the locus of points traced
out by the ``measured'' values from individual simulations can be
easily approximated by simply plotting curves that intersect the
entire swarm of simulated points. Doing this we find that $N / N_{mdp}
= \beta_{obs,1d}^2 / 9.2$ (lower dashed curve in Figure 9), which
agrees with the result deduced above from equation 10.

We also explored simulations where the true amplitude $a_0$ was set
equal to the MDA. Figure 10 shows the cumulative distribution of the
``measured" values of $\beta$ from several such simulations.  One can
see from Figure 10 that roughly 60\% of the time one would ``measure"
the amplitude, $a$, at the 3$\sigma$ level, or better.  We emphasize
that this relation is based on the 1-d (single parameter) confidence
range for $a$, and would be appropriate only for the case of
addressing the question of the detection of a significant modulation
amplitude independent of the position angle.  We now explore the
joint, 2-dimensional confidence regions.

\subsection{Joint Confidence Regions for $a$ and $\phi$}

For a 2-dimensional confidence region we need to find the contour in
the $a$ - $\phi$ plane that encloses a specified fraction of the
best-fit pairs.  For a $1\sigma$ region (68.3\% confidence) this is
the contour that satisfies $\Delta\chi^2 (a, \phi) = 2.3$, where
$\Delta\chi^2 = \chi^2 (a, \phi) - \chi^2_{min} (a_{best},
\phi_{best})$.  We again use the Stokes decomposition and compute
$\Delta\chi^2 (I, Q, U)$ on grids of $I$, $Q$ and $U$ around the
best-fit values, $I_{best}$, $Q_{best}$, and $U_{best}$. We can then
convert the grids of Stokes parameters into the appropriate values of
$a$ and $\phi$ and find the boundaries of the region that satisfies
$\Delta\chi^2 \le 2.3$. We next find the maximum and minimum value on
the boundary for each parameter.  We can then define $\sigma_{a,2d} =
(a_{max} - a_{min})/2$ and $\sigma_{\phi} = (\phi_{max} -
\phi_{min})/2$, where $a_{max}$, $a_{min}$, $\phi_{max}$ and
$\phi_{min}$ are the maximum and minimum values on the contour of $a$
and $\phi$, respectively.  Figure 11 shows a pair of $1\sigma$
confidence regions computed in this fashion.  Results from two
simulations are shown, one with a lower amplitude, $a_{0,low} = 2/24$,
and one with a higher amplitude, $a_{0,hi} = 3/25$. Both simulations
were performed with $\phi_0 = 25$ deg (these true parameters are
marked by the red diamond symbols), and N = 10000.  In each case a
modulation curve was randomly sampled using the true amplitude and
position angle, and the simulated data were then fitted to determine
the best-fit values of the amplitude and position angle.  These points
are shown by the green square symbols.  The shaded areas show the
regions of $a$ and $\phi$ around each best-fit pair that satisfy
$\Delta\chi^2 \le 2.3$. The horizontal and vertical dotted lines mark
the maximum and minimum values on the regions for each parameter.  One
can see that the size of the confidence region grows for smaller
intrinsic modulation amplitudes (as one would expect), and that the
confidence regions are in general not circular. The fact that the best
fit value (green square symbol) and $1\sigma$ contours did not enclose
the ``true'' values (red diamond symbol) for the higher amplitude
example was just the ``luck of the draw'' for this single realization.

The probability distribution expressed earlier (\S 3) in equation (7)
can also be used to derive an analytic expression for the 2-d
confidence contour in the $a$ - $\phi$ plane for any desired level of
confidence, as long as the intrinsic amplitude is not very large (as
is satisfied by the simulations just presented). This approach has
been investigated by Weisskopf et al. (2010). They derive a pair of
parametric relations for the values of $a$ and $\phi$ on any
confidence contour (see their equation 8). These expressions predict
the correct range (extremes) in the amplitude, $a$, but overpredict
the range in $\phi$ by a factor of 2, apparently because their
original derivation neglected a switch from phase angle to position
angle (Weisskopf, private communication). Thus, if one replaces all
occurrences of $\phi$, $\phi_0$ and $\psi$ with $2{\rm x}$ the
respective angle (eg., $\sin\phi_0 --> \sin 2\phi_0$) beginning at
their equation (6), one obtains the following expressions;
\begin{equation}
a =\left ( a_0^2 + \Delta a_C^2 + 2a_0\Delta a_C \cos 2(\psi - \phi_0) 
\right )^{1/2}
\end{equation}
and 
\begin{equation}
\phi = 1/2 \tan^{-1} \left ( (a_0\sin 2\phi_0 + \Delta a_C \sin 2\psi ) 
/ (a_0\cos 2\phi_0 + \Delta a_C \cos 2\psi ) \right )  , 
\end{equation}
where $\Delta a_C = (-4 \ln (1 - C) / N )^{1/2}$, with $C$ and $N$
being the desired confidence level and number of detected photons,
respectively, and $\psi$ is just a parametric angle that varies around
the contour.  The thick blue curves in Figure 11 were drawn using these
expressions with $N = 10,000$, $C = 0.683$ (ie., $1\sigma$), and the
pair $(a_0, \phi_0)$ given by the appropriate best fit values (green
square symbols).  These contour curves provide an excellent match to
the boundaries of the shaded regions, as we would expect at these
relatively low amplitudes for which equation (7) is a good
approximation to the exact probability density.

We can now compute the value $\beta_{2d} = a_{best} / \sigma_{a,2d}$
for each particular simulation.  This again quantifies the ``number of
sigmas'' of the measurement, but now reflects the fact that it is a
{\it joint} measurement of both $a$ and $\phi$ together.  Results from
a number of such simulations are also shown in Figure 9, where we plot
the same figure of merit, $N / N_{mdp}$, as before, but now using
$\beta = a_{best} / \sigma_{a,2d}$.  This curve is again quadratic in
$\beta$ but rises more steeply than the 1d relation, because
$\sigma_{a,2d}$ is larger than $\sigma_{a,1d}$.  In agreement with
Weisskopf et al. (2010) we find that the 2d relation is very well
approximated as $N / N_{mdp} = \beta^2 / 4.1$, and this is the dashed
line running through the square symbols. 

To further quantify the behavior as a function of amplitude we
explored additional simulations with larger modulation amplitudes. An
example of how the confidence contours defined by equations (12) and
(13) (which were derived from equation (7)) over-predict the true size
at large amplitudes is shown in Figure 12. The symbols have the same
meanings as in Figure 11. This simulation had $a_0 = 2/3$ and $\phi_0
= 30 \; ({ \rm deg})$, and was computed with $N = 8000$ events. The
shaded region is again the joint $1\sigma$ confidence region
satisfying $\Delta\chi^2 \le 2.3$, and the blue curve, which clearly
over-predicts the confidence region, is from equations (12) and (13).
For completeness we also determined $N / N_{mdp}$ for several larger
amplitude values.  We find that for amplitudes $a_0 = 1/3$, $1/2$,
and $2/3$, $N / N_{mdp} \approx \beta^2 / 4.4$, $\beta^2 / 5.1$, and
$\beta^2 / 6.1$, respectively.

We also show in Figure 13 the position angle uncertainty,
$\sigma_{\phi}$ in degrees (at $1\sigma$ confidence) as a function of
$\beta = a_{best} / \sigma_{a,2d}$.  We find that this relation can be
very well approximated as $\sigma_{\phi} = 28.5 ({\rm deg}) / \beta $.
The $\beta^{-1}$ dependence can be seen starting with equation 11.
With $\sigma_Q = \sigma_U \equiv \sigma_{Q,U}$ and a bit of algebra
one can show that $\sigma_{\phi}^2 = (180/\pi)^2 \sigma_{Q,U}^2 / 4
(U^2 + Q^2)$.  Substituting for $\sigma_{Q,U}^2 \approx 2N/M^2$ and
replacing $(Q^2 + U^2) = a^2 I^2$ we find that $\sigma_{\phi} \propto
(1/a)(1/N^{1/2})$.  A final substitution of $a = a_{MDP} = 4.29 /
N_{MDP}^{1/2}$ shows that $\sigma_{\phi} \propto (N/N_{MDP})^{-1/2}$.
Since we previously demonstrated that $N/N_{MDP}$ scales like
$\beta^2$, this shows that $\sigma_{\phi} \propto \beta^{-1}$.  Thus,
a joint $3\sigma$ measurement constrains the position angle to better
than 10 degrees.

\section{Discussion and Summary}

The MDP is a very commonly employed figure of merit to describe the
polarization sensitivity of a detector system. In their recent paper
Weisskopf et al. (2010) have argued that ``more counts would be
needed" to {\it measure} the polarization corresponding to the $99\%$
confidence MDA rather than to just establish the same level.  As we
have shown this is correct in the sense of a ``polarization
measurement'' being a joint measurement of both the amplitude $a$ and
position angle $\phi$.  This seems sensible, since it adds an
additional requirement, that the measured $a$ and $\phi$ both fall
within a 2d confidence region.  We have also shown that in the limit
of small amplitudes the additional number of counts required is given
by a factor of $\approx 2.2$.  This factor decreases slowly with
increasing intrinsic amplitude. For example, it has a value of
$\approx 1.5$ when $a_0 = 2/3$.

However, if one were to ask a different question, and were interested
in establishing simply that a source is polarized, with say, a
$3\sigma$ measurement of the amplitude, $a$, without concern for the
position angle $\phi$, then the 1d (one interesting parameter) case is
appropriate, and no extra counts are needed to measure the MDA.  So,
in some sense the answer one obtains depends on the question asked.

The MDP value has often been used to estimate observing times required
to reach particular sensitivity levels. The results shown here
demonstrate that exposure requests should clearly match the
measurement goals of the desired scientific program.  For example, if
source modeling requires a joint measurement of both polarization
parameters, then the appropriate time to reach a required precision
for two interesting parameters should be requested.

\acknowledgments

It is a pleasure to thank Jean Swank, Keith Jahoda, Kevin Black, Joe
Hill, Craig Markwardt, Phil Kaaret and Charles Montgomery for helpful
discussions related to X-ray polarization measurements and
instrumentation.  We also thank Martin Weisskopf, Ronald Elsner and
Stephen O'Dell for reviewing an earlier version of this manuscript,
and much helpful feedback on the question of measuring polarization in
the X-ray band.  Finally, we thank the anonymous referee for their
careful review which helped us improve this paper.

\clearpage



\begin{figure}
\epsscale{0.75}
\plotone{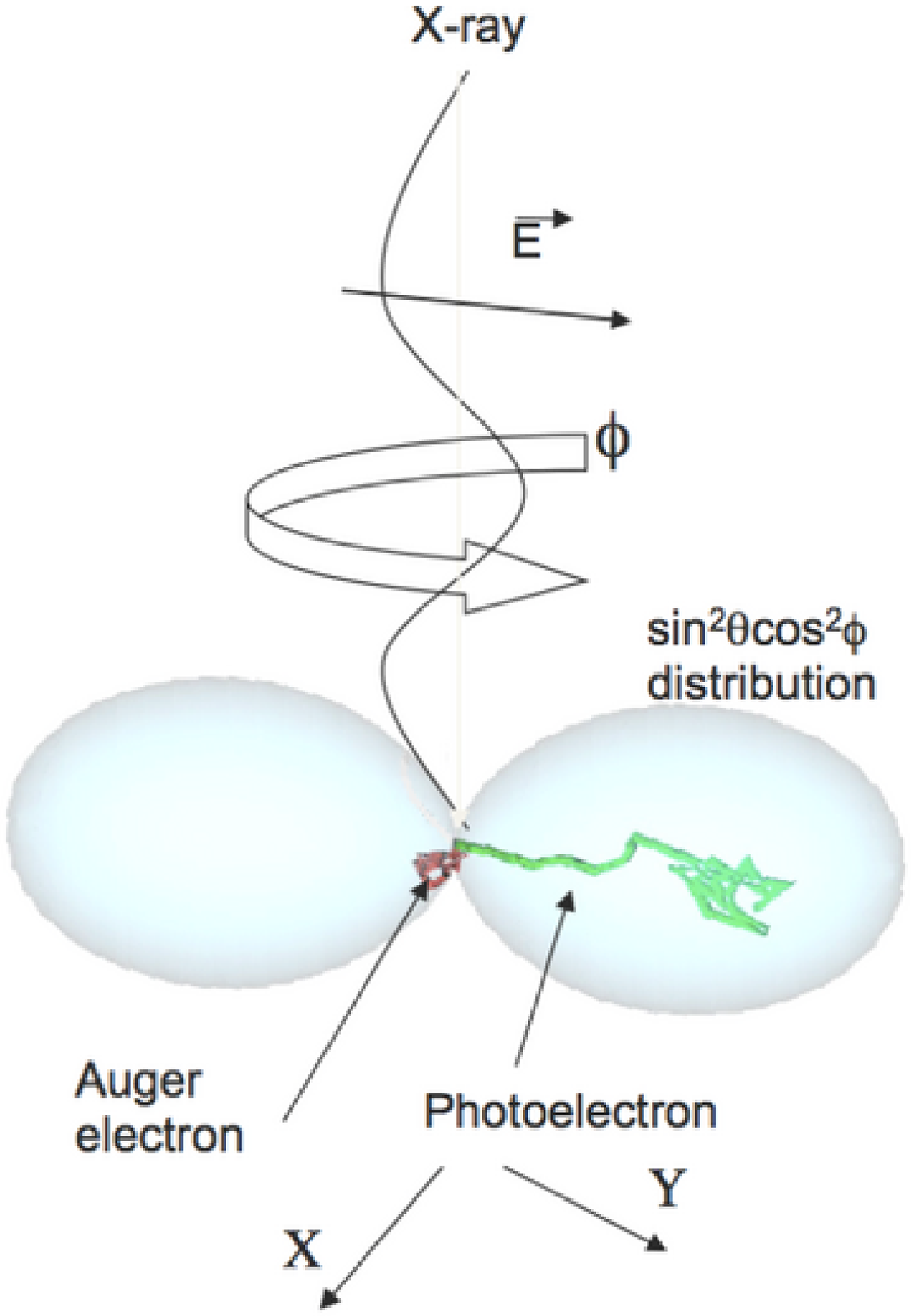}
\caption{Geometry relevant for photoelectric polarimeter
measurements. Photons travel from figure top to bottom. Photoelectrons
are preferentially emitted in the plane perpendicular to the photon
direction of travel (in this case the X - Y plane), and their initial
direction is indicated by the azimuthal angle $\phi$.  \label{fig1}}
\end{figure}

\clearpage

\begin{figure}
\epsscale{0.75}
\plotone{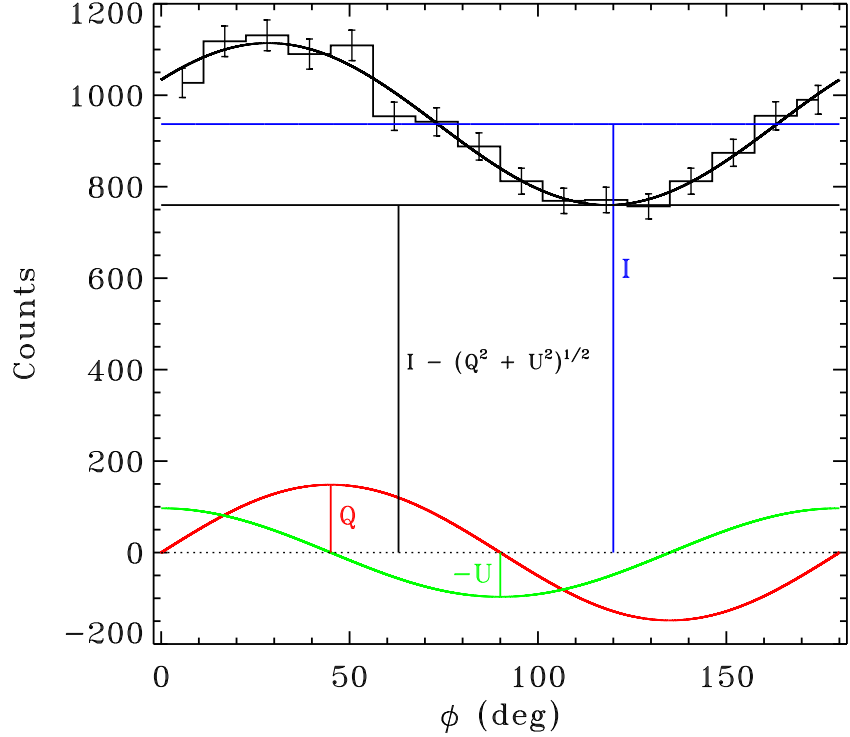}
\caption{An example of a simulated modulation curve (histogram with
error bars) and best fitting model (smooth black curve).  This example
is a single statistical realization of 15,000 events from the
distribution $S(\phi) = 10 + 5 cos^2 (\phi - 30 (deg))$. The three
best-fit Stokes components, $I$ (blue), $Q$ (red), and $U$ (green) are
also indicated.  The solid black curve is the sum of the colored
curves. The unpolarized level, $I - (Q^2 + U^2)^{1/2}$, is also
indicated. See the text (\S 4) for further details of this particular
simulation. \label{fig2}}
\end{figure}

\clearpage

\begin{figure}
\epsscale{0.75}
\plotone{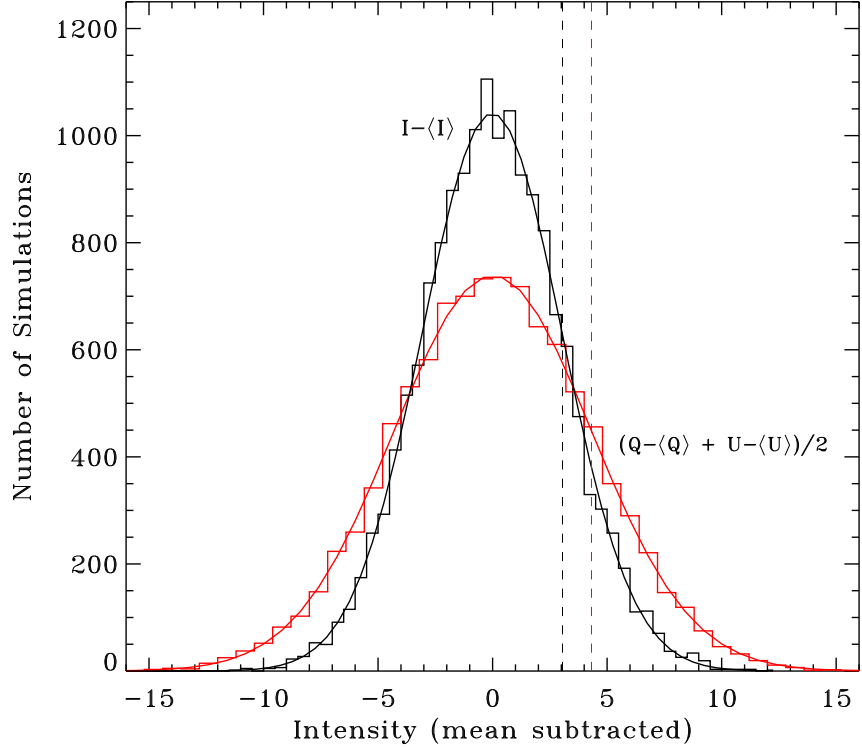}
\caption{Example distributions of the fitted Stokes parameters $I$,
$Q$ and $U$ from 10,000 simulations of the same angular distribution,
$S(\phi)$, shown in Figure 2.  For each simulation a modulation curve
was computed with $M=40$ phase angle bins.  Shown are the
mean-subtracted distributions of $I$ and $(Q + U)/2$ along with the
best-fitting Gaussian functions.  The standard deviations determined
from the fits are indicated by vertical dashed lines, and are
consistent with $\sigma_I = (15,000/(40^2))^{1/2}$ and $\sigma_{Q,U}=
(2*15,000/(40^2))^{1/2}$.
\label{fig3}}
\end{figure}

\clearpage

\begin{figure}
\epsscale{0.75}
\plotone{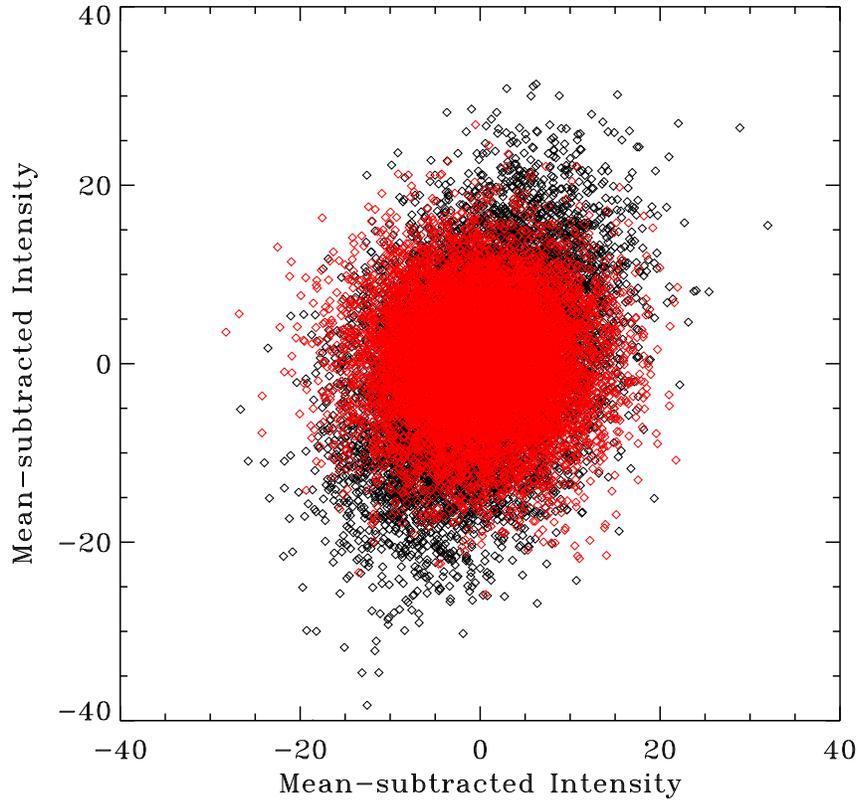}
\caption{Distributions of $I - <I>$ versus $Q - <Q>$ (black diamonds)
and $Q - <Q>$ versus $U - <U>$ (red diamonds) from a representative
simulation with a large intrinsic amplitude of 16/30. The black
diamonds show a clear correlation whereas the red symbols are
circularly symmetric (no evident correlation). See \S 4 for further
details of this particular simulation.
\label{fig4}}
\end{figure}

\clearpage

\begin{figure}
\epsscale{0.85}
\plotone{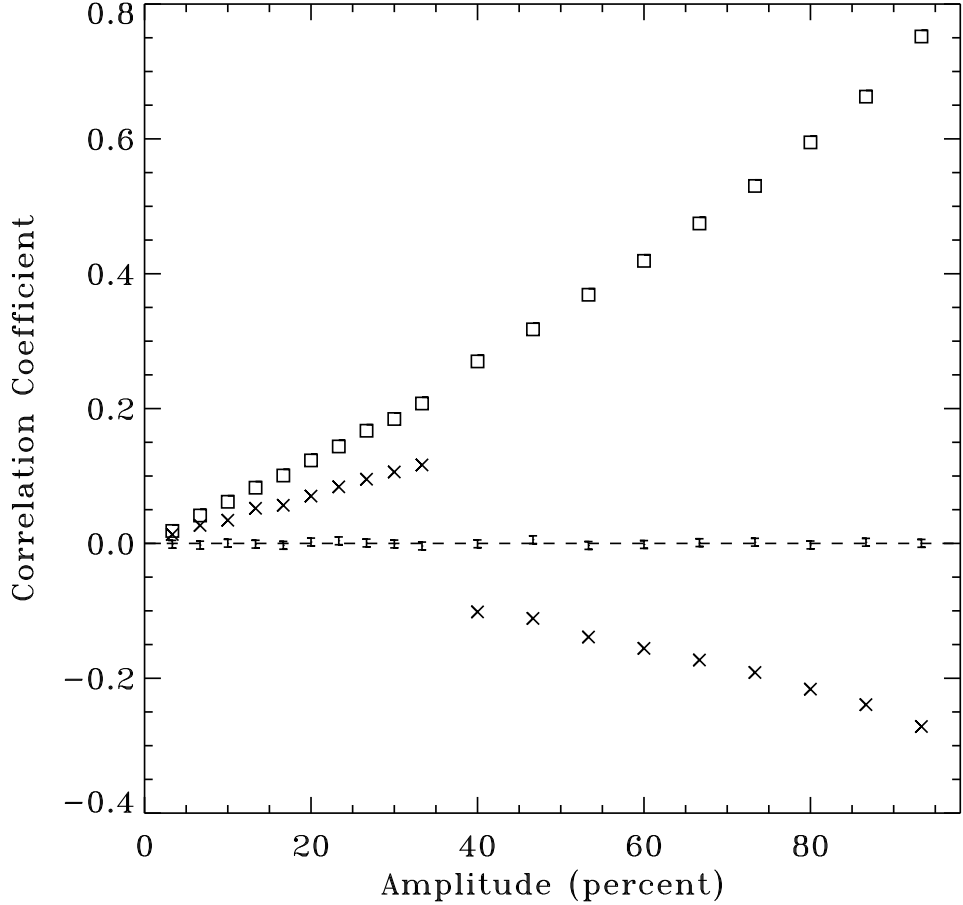}
\caption{Linear correlation coefficients as a function of intrinsic
amplitude for a representative set of simulations. The coefficients
for $Q$ versus $U$, $Q$ versus $I$, and $U$ versus $I$ are denoted by
the ``error bar,'' ``square,'' and ``$\times$'' symbols, respectively.
These results were computed from 100,000 realizations. The error bars
have a size of $(1/100,000)^{1/2}$, and are only plotted for the $Q$
versus $U$ results but are appropriate to all the coefficients.  The
simulations with $a < 35\%$ had $\phi_0 = 30 ({\rm deg})$, while
those with $a > 35\%$ had $\phi_0 = 55 ({\rm deg})$. This accounts
for the ``discontinuous'' behavior of the $Q$ versus $I$ and $U$
versus $I$ coefficients. This plot clearly demonstrates that $Q$ and
$U$ are statistically independent variables, but that $Q$ and $U$ are
both correlated with $I$ in an amplitude dependent manner.
\label{fig5}}
\end{figure}

\clearpage

\begin{figure}
\epsscale{0.85}
\plotone{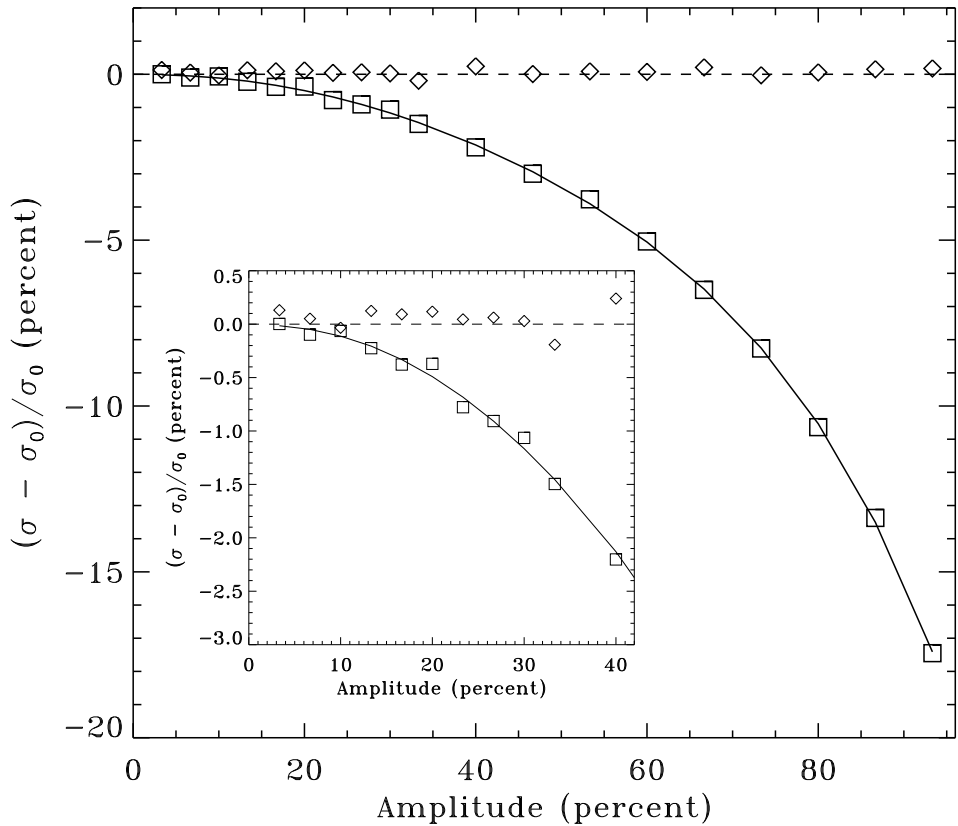}
\caption{Standard deviations of the distributions in $I$, $Q$ and $U$
versus amplitude from a large suite of simulations. The results are
shown as the percentage deviation from $\sigma_0$, the expected
variance in the limit that equation (7) is valid.  Shown are the
results for the distribution of $I$ (diamond symbols) and, since the
distributions of $Q$ and $U$ have the same variance, the average for
the $Q$ and $U$ standard deviations (square symbols) from 400,000
realizations at each amplitude value.  The solid curve is a 4$^{th}$
order polynomial fit to the $Q$ and $U$ values. The inset panel shows
an expanded view of the low amplitude behavior. See \S 4 for futher
details of these simulations.
\label{fig6}}
\end{figure}

\clearpage

\begin{figure}
\epsscale{1.0} \plotone{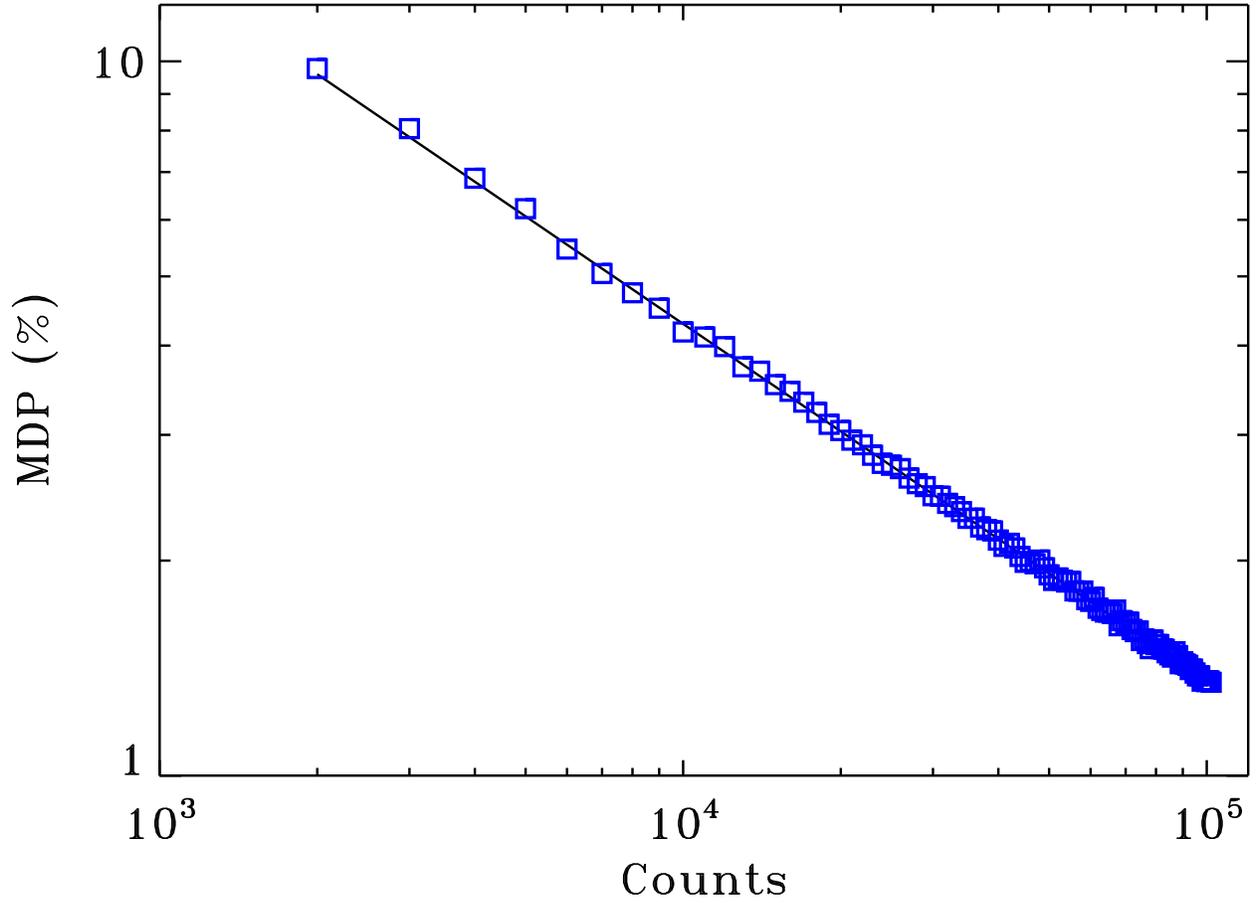}
\caption{The amplitude, $a_{1\%} \equiv MDP \; (\mu = 1)$, vs the number
of counts, $N$, obtained from simulations described in \S 4.1 (blue
squares), and the analytic formula (solid curve, \S 3). \label{fig7}}
\end{figure}

\clearpage


\begin{figure}
\plottwo{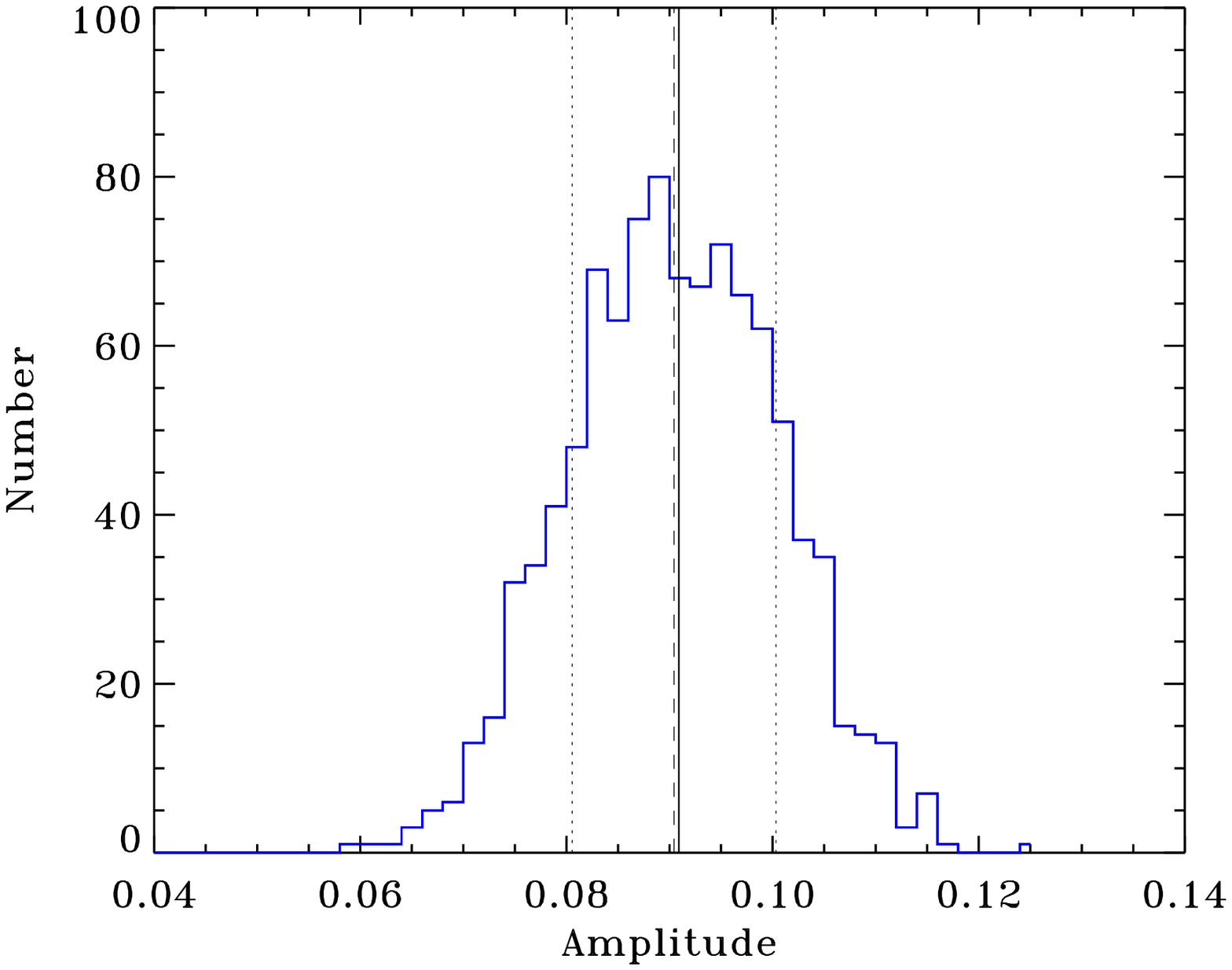}{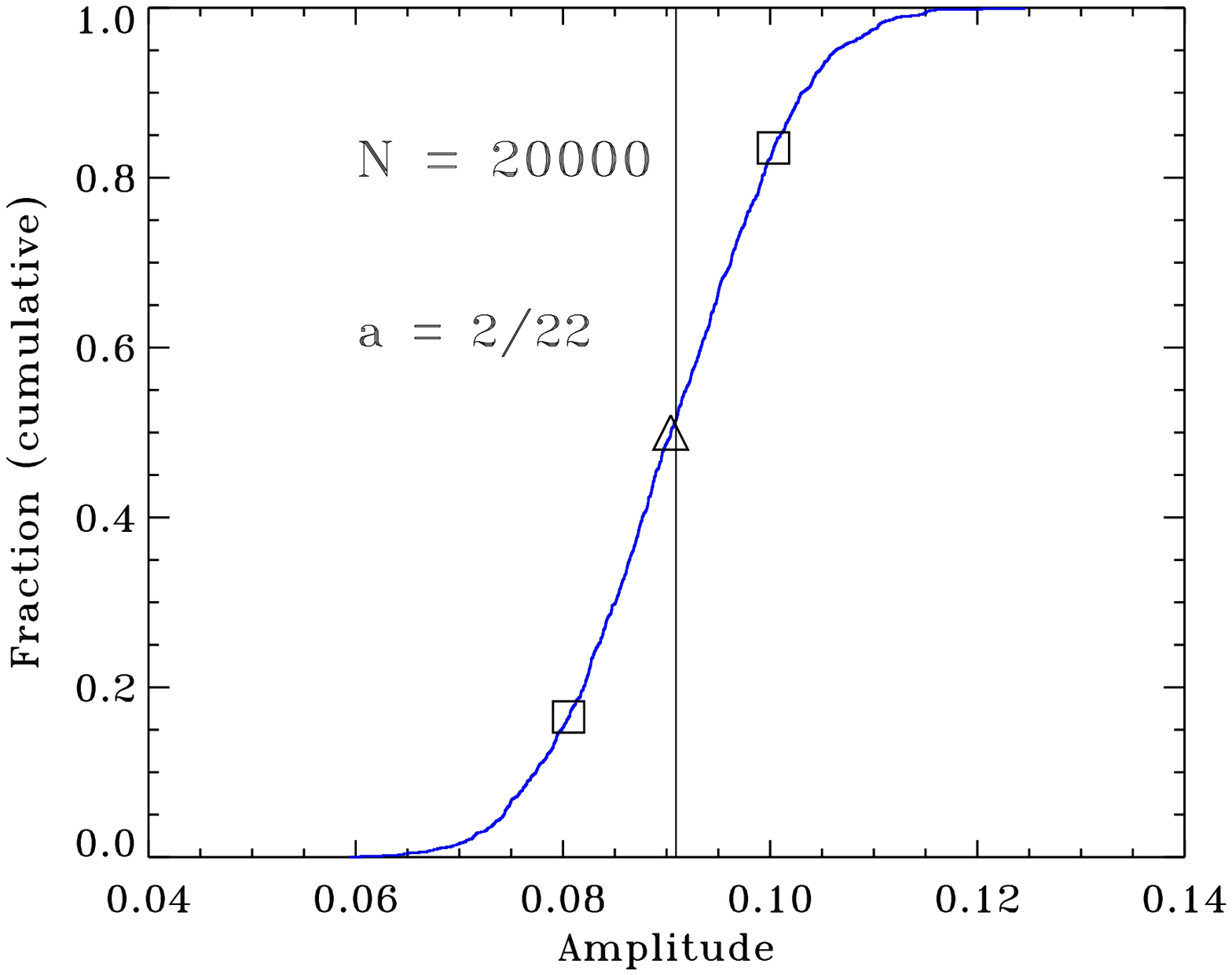}
\caption{Example 1d distributions in the amplitude, $a$, computed from
Monte Carlo simulations. The left panel shows the differential
distribution (binned), and the right panel shows the cumulative
distribution. To estimate the mean we find the midpoint (triangle
symbol), and the $1\sigma$ extremes, $a_{min}$ and $a_{max}$ at
cumulative fractions of 0.16 and 0.84, respectively (square symbols,
enclosing 68\% of the distribution). The corresponding amplitude
values are also indicated with vertical dottend lines in the left
panel. \label{fig8}}
\end{figure}

\clearpage

\begin{figure}
\epsscale{1.0}
\plotone{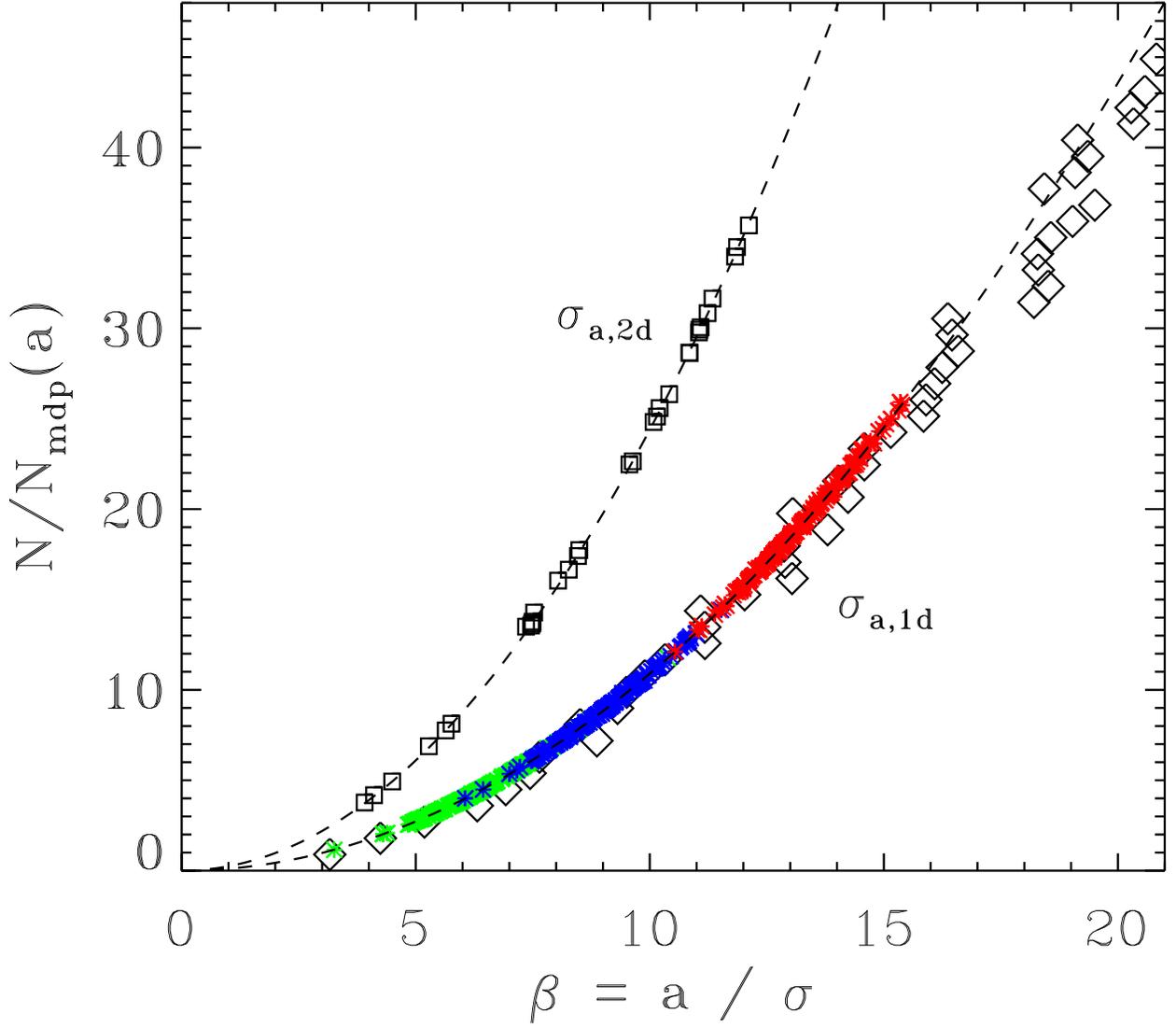}
\caption{Plot of $N / N_{mdp} (a)$ as a function of $\beta$, the
``number of sigmas'' of the measurement. Both the 1d amplitude
distribution ($\sigma_{a,1d}$, independent of the position angle) and
the 2d joint distribution ($\sigma_{a,2d}$), are shown. The black
diamond symbols are derived from the results of $M_{sim} = 1000$
simulations for different values of $N$ (see discussion in \S 4.3).
The colored symbols are the results of individual simulations where
$\beta$ is derived from the best-fit amplitude, $a$, and its 1d,
$1\sigma$ uncertainty, $\sigma_{a,1d}$.  The red symbols were computed
with $N=10,000$, $a_0 = 2/22$, $\phi_0=0$ (deg), the blue used
$N=20,000$, $a_0 = 2/22$, $\phi_0=0 (deg)$, and the green with
$N=24,000$, $a_0 = 3/25$, and $\phi_0=22.5$ (deg).  The solid dashed
curve for the 1d case is given by $N / N_{mdp} (a) = \beta^2 /
9.2$. The black square symbols show the 2d confidence region results
(see discussion in \S 4.4). The curve running through the 2d results
is given by $N / N_{mdp} (a) = \beta^2 / 4.1 $\label{fig9}}
\end{figure}

\clearpage

\begin{figure}
\epsscale{1.0}
\plotone{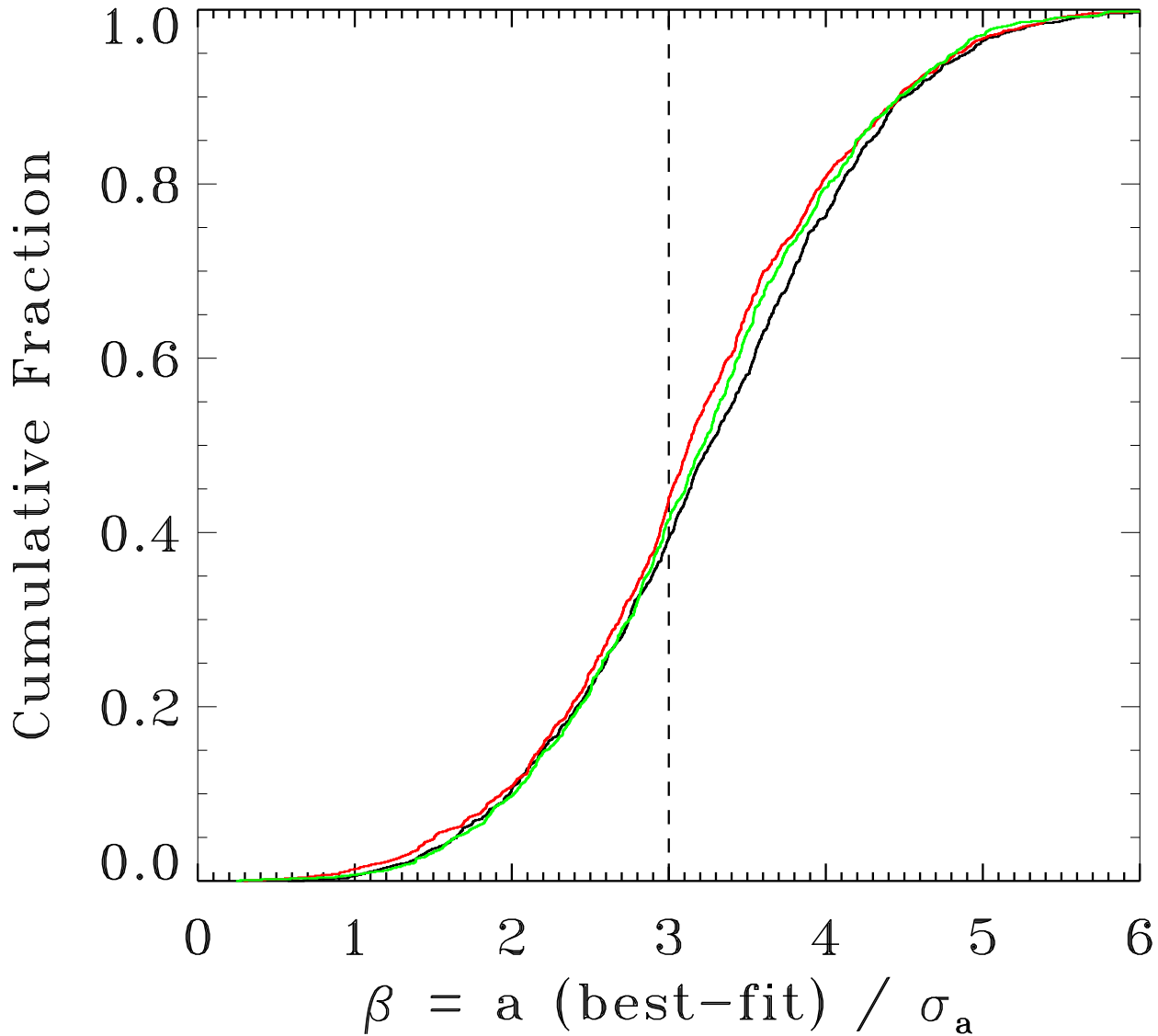}
\caption{Plot of the cumulative distribution of $\beta = a /
\sigma_{a,1d}$ for several different simulations all satisfying the
condition that $a =$ MDA. A vertical line is plotted at $\beta = 3$
(the nominal $3\sigma$ detection criterion), and which is close to the
most probable value (the distribution is not exactly gaussian,
ie. symmetric). See the discussion in \S 4.3 for more
details. \label{fig10}}
\end{figure}

\clearpage

\begin{figure}
\epsscale{1.0}
\plotone{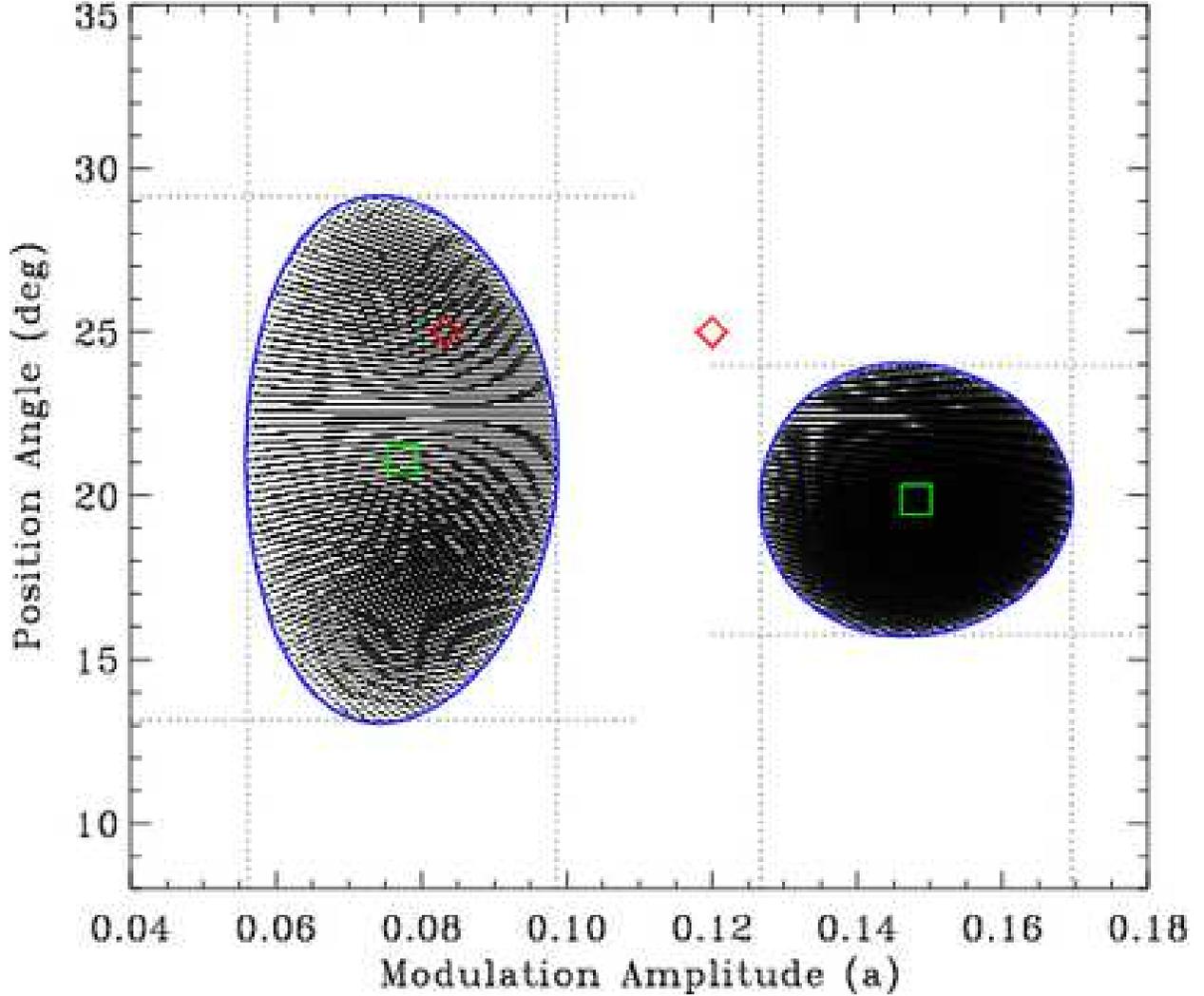}
\caption{Confidence regions and contours in the (a, $\phi$)
plane. Results from two simulations are shown. Simulated modulation
curves were computed with N=10,000 counts for two different intrinsic
amplitudes both with a position angle of 25 degrees. These values are
marked by the red diamond symbols. The resulting best fit parameter
values (green squares) and confidence regions (shaded areas) are
shown. The shaded regions are the $\Delta \chi^2 < 2.3$ regions
($1\sigma$ for 2d confidence regions). The horizontal and vertical
dotted lines denote the extremes in each parameter and are used to
compute $\sigma_{a,2d}$ and $\sigma_{\phi ,2d}$. The blue contour curves
were computed from the analytical expressions in \S 4.4.
\label{fig11}}
\end{figure}

\clearpage

\begin{figure}
\epsscale{1.0}
\plotone{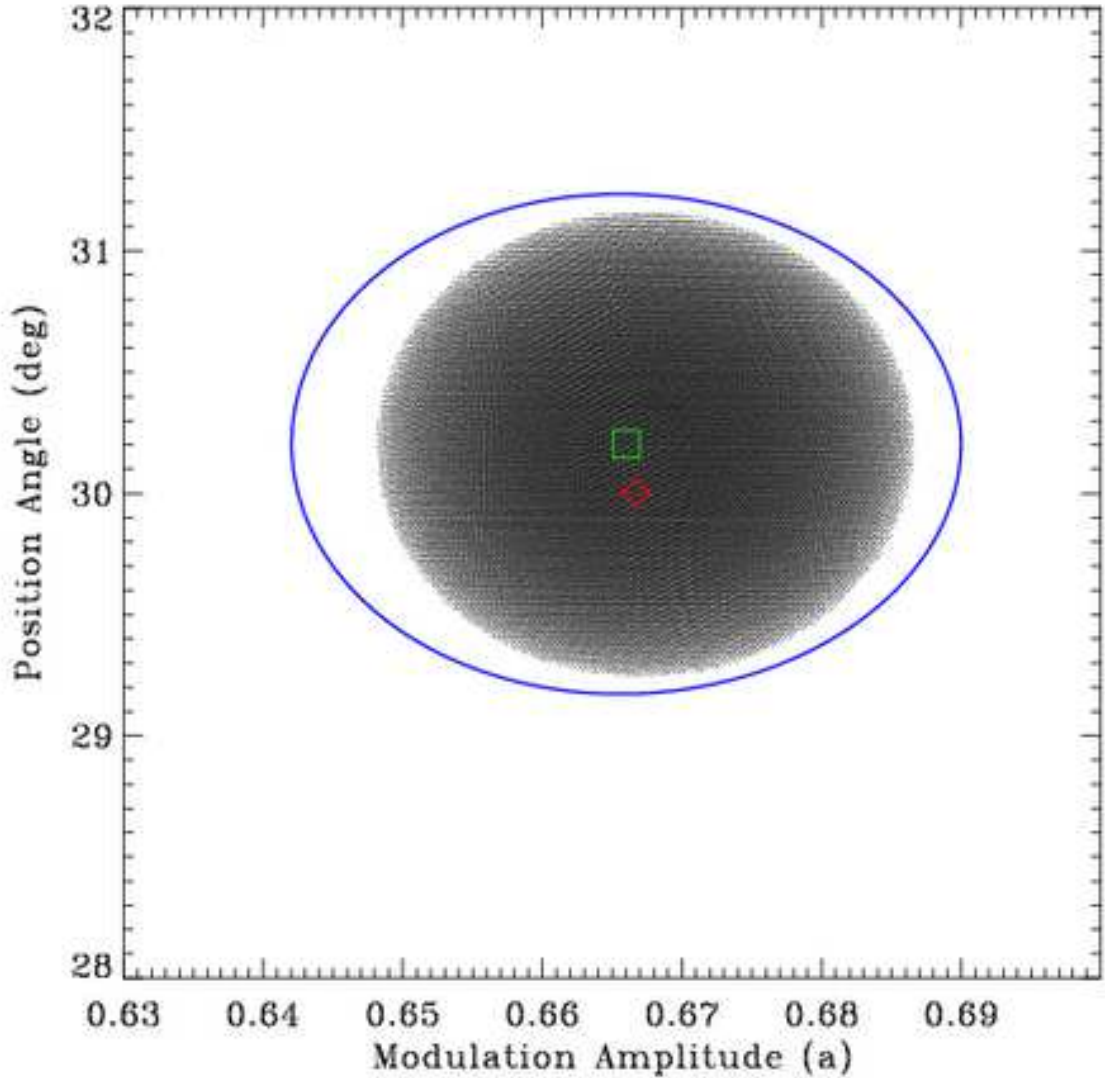}
\caption{Confidence region in the (a, $\phi$) plane from a simulation
with a large intrinsic amplitude $a_0 = 2/3$. Symbols are the same as
in Figure 11.  The shaded area is the $\Delta \chi^2 < 2.3$ region
($1\sigma$ for 2d confidence regions). The blue contour curve was
computed from the analytical expressions in \S 4.4 (equations 12 and
13), and clearly overpredicts the size of the confidence region for
this large amplitude. See the text in \S 4.4 for further details.
\label{fig12}}
\end{figure}

\clearpage

\begin{figure}
\epsscale{1.0}
\plotone{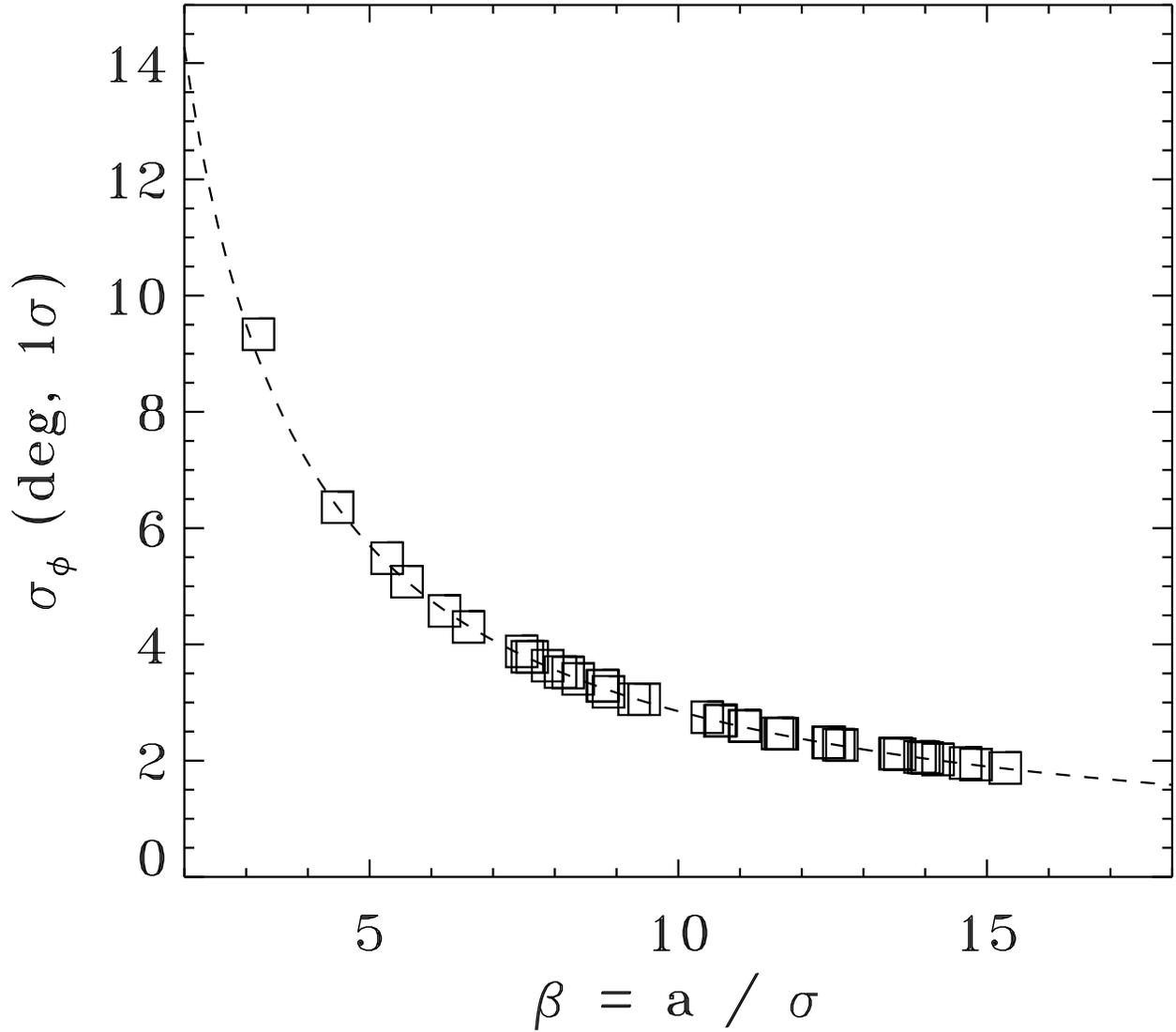}
\caption{Plot of the position angle uncertainty, $\sigma_{\phi}$
($1\sigma$, in degrees), derived from the 2d confidence regions, as a
function of $\beta_{2d}$, the ``number of sigmas'' of the measurement.
The solid dashed curve is given by $\sigma_{\phi} = 28.5 /
\beta_{2d}$. 
\label{fig13}}
\end{figure}

\clearpage

\end{document}